\documentclass[draft]{agujournal2019}
\usepackage{url} 
\usepackage{lineno}
\usepackage{graphicx}
\usepackage{mathtools}
\usepackage{mathrsfs}
\usepackage{amssymb}
\usepackage{amsmath}
\usepackage{amsfonts}
\usepackage{bm}
\usepackage{tabu}
\usepackage{makecell}
\usepackage[inline]{trackchanges} 
\usepackage{soul}
\usepackage{lmodern} 
\usepackage{CJKutf8}

\usepackage[OT2,T1]{fontenc}
\newcommand\textcyr[1]{{\fontencoding{OT2}\selectfont #1}}
\journalname{Geophysical Research Letters}

\begin{document}

%
%


\title{A strictly geostrophic product of sea-surface velocities from the SWOT fast-sampling phase}

%
%




\authors{Takaya Uchida\,(\begin{CJK}{UTF8}{min}内田貴也\end{CJK})\affil{1,2}\thanks{Now at \textit{Climate Dynamics Laboratory, Center for Earth Sciences, Moscow Institute of Physics and Technology (\textcyr{МФТИ}), Dolgoprudny, Russia}.},
Badarvada Yadidya\affil{1,3,4},
Vadim Bertrand\affil{2},
Jia-Xuan~(Julia)~Chang\affil{4},
Brian K.~Arbic\affil{4},
Jay F.~Shriver\affil{3},
and Julien Le Sommer\affil{2}
}


\affiliation{1}{Center for Ocean-Atmospheric Prediction Studies, Florida State University, Florida, USA}
\affiliation{2}{Universit\'e Grenoble Alpes, CNRS, INRAE, IRD, Grenoble INP, Institut des G\'eosciences de l’Environnement, Grenoble, France}
\affiliation{3}{Naval Research Laboratory, Stennis Space Center, Mississippi, USA}
\affiliation{4}{Department of Earth and Environmental Sciences, University of Michigan, Michigan, USA}




\correspondingauthor{T.~Uchida}{takachanbo@gmail.com}



\begin{keypoints}
\item We apply dynamic mode decomposition (DMD) to extract the strictly geostrophic component from the global SWOT one-day-repeat orbit.
\item Frequency-wavenumber spectra of sea-surface height anomaly (SSHa) display distinctive features between the geo- and ageo-strophic component.
\item Vorticity-strain joint probability density functions (PDFs) show that our geostrophic velocities largely satisfy small Rossby numbers.
\end{keypoints}

%
%

%
%


\begin{abstract}
While geostrophy remains the simplest and most practical balance to extract velocity information from sea-surface height anomaly (SSHa), confusions remain within the oceanographic community to what extent this balance can be applied to altimetric observations with the launch of the Surface Water and Ocean Topography (SWOT) satellite. Given the limited temporal resolution of SWOT, many studies have resorted to claiming that the spatially filtered SSHa fields correspond to the geostrophic component. This introduces the ambiguity of which spatial scale to choose. Here, we build upon the recent developments in internal tide (IT) corrections \cite{yadidya2025predicting} and apply a dynamic mode decomposition (DMD)-based method introduced by \citeA{lapo2024multi} to robustly extract the geostrophic component associated with sub-inertial frequencies from the SWOT one-day-repeat orbit; we distribute the global dataset as a public good. We provide the joint probability density function (PDF) of vorticity and strain, and spectra of SSHa at a few cross-over regions.
\end{abstract}

\section*{Plain Language Summary}
Information on how the ocean flows, namely velocity, is crucial to study the physics of the ocean and to quantify the transport of matter by currents. Satellites now routinely observe the up-and-downheavals of sea-surface elevation. With the launch of the SWOT satellite, we can now globally monitor the elevations at the scales of two kilometers. 
The elevation signals are a superposition of waves and ocean currents.
We apply a mathematical method, mrCOSTS (pronounced \textit{Mister-Kosts}), to extract spatial patterns of sea-surface elevation that are in balance with the ocean currents.
We show that such patterns evolve on scales larger than hundreds of kilometers in space and slower than tens of days in time.

%
%

%


%
%
%
%

\section{Introduction}
Geostrophy is a dynamical balance between the Coriolis inertial and horizontal pressure gradient forces that emerges from the rotating Navier-Stokes equation under a Taylor expansion for small Rossby numbers \cite{phillips1963geostrophic,gill2016atmosphere,pedlosky1984equations,Vallis:2006aa,early2025measuring}. 
What this approximation entails is that geostrophic balance should only apply to signals that evolve on time scales longer than the inertial period and spatial scales larger than the internal Rossby radii of deformation \cite{Torres2018,Wang_2023,wang2023simple}.
It is common, however, for studies that venture into the real ocean to play it loose and fast about this premise. Particularly for studies that adopt data from SWOT observations \cite{dibarboure2024blend,dibarboure2025blending}, the constraint on time scales is often overlooked, partially attributable to the limited sampling frequency of SWOT, and they tend to settle with an ad-hoc spatial filtering \cite<e.g.,>[]{carli2023ocean,carli2025southern,wang2025cross,qiu2025fine,villas2025observing,tchonang2025swot}.
This introduces an arbitrary dependency on what spatial scale to choose for the filter, inconsistencies between the theory and application of geostrophy, and potential errors in the estimation of velocities \cite{yu2021geostrophy,uchida2025dynamic,zhang2025swot}.

In order to address this issue, we produce and publicly distribute a global dataset of strictly geostrophic velocities from the fast-sampling SWOT Calibration and Validation (Cal/Val) orbit; we extract SSHa signals that satisfy small Rossby number arguments in both space-and-time dimensions. The method by which we achieve this is DMD, a linear-algebraic method akin to frequency-wavenumber spectral decomposition but negates the necessity of periodic boundary conditions \cite{kutz2016dynamic,lapo2025phasor}.
Conceptually, it can be thought of as empirical orthogonal function (EOF) spatial modes where each mode is tied to a specific temporal frequency and phase \cite{towne2018spectral}.

DMD has its history in the field of fluid mechanics \cite{schmid2022dynamic,baddoo2023physics,linot2025extracting} but in the context of oceanography, it has been applied to capture tidal patterns across the Strait of Gibraltar \cite{dias2025analysis}, identify spatiotemporal structures in global sea-surface temperature such as the El Ni\~no Southern Oscillation \cite{kutz2016multiresolution,Franzke_2022}, separate waves from turbulence \cite{chavez2024wave}, and develop stochastic eddy parametrizations \cite{li2023dmd,tucciarone2025derivation}.

In the following, we briefly describe the dataset and methodology.
Results are given in Section~3 and we conclude in Section~4.

\section{Data and Methods}
\subsection{SWOT data and mrCOSTS}
\label{sec:data}
We take the Level 3 (L3) SWOT data from its fast-sampling phase (from March 29$^\text{th}$ to July 11$^\text{th}$,~2023) where IT imprints on SSHa are removed using the HYbrid Coordinate Ocean Model (HYCOM) operational forecast produced by the U.S. Navy \cite{chassignet2009us}.
We shall refer to this de-tided dataset as L3$_{\text{HYCOM}}$.
This has the advantage, compared to the High Resolution Empirical Tides (HRET) model used in the default SWOT product \cite<>[Section 5.9 of their handbook, \url{https://www.aviso.altimetry.fr/fileadmin/documents/data/tools/handbook_duacs_SWOT_L3.pdf}]{swot2024}, of removing not only the signals of coherent ITs but also incoherent ITs \cite{yadidya2024phase,yadidya2025predicting}.
We shall exclude $\pm5^\circ$ about the Equator where the inertial frequency becomes small ($f\rightarrow 0$) and grid points that have data for less than half of the time from our analyses; the former excludes the equatorial region where time scales are weakly constrained by the inertial frequency and small Rossby number arguments are ill defined, and the latter removes polar regions where coverage varies with sea-ice extent.

\citeA{uchida2025dynamic} recently demonstrated the skill of multi-resolution COherent Spatio-Temporal scale Separation (mrCOSTS) in extracting the geostrophic component from SSHa in the separated Gulf Stream region.
Here, we extend their study and apply mrCOSTS over the entire SWOT fast-sampling phase. 
A notable update is that ITs are removed using the HYCOM forecast instead of HRET.
MrCOSTS is a variant of DMD; it is particularly versatile to data comprising of various frequencies by recursively applying DMD \cite{lapo2024multi}.
It has also been applied to identify spatiotemporally coherent structures in the atmospheric boundary layer \cite{lapo2025scale}.
Details on the methodology are left to \citeA<>[schematic in their Fig.~1]{lapo2024multi} and \citeA{uchida2025dynamic} but we briefly describe the procedural steps taken for each SWOT L3 pass: 
\begin{enumerate}
    \item HRET contributions are added back and then IT signals are removed using the HYCOM forecast (L3$_{\text{HYCOM}}$).
    \item Missing data due to poor quality flags were spatially interpolated over bi-linearly. 
    \item An isotropic Gaussian spatial filter with a standard deviation of three grid points ($\sim 6$\,km) was applied. 
    \item When more than 30\% of the swath had data missing, that day was dropped and linearly interpolated over in time. Islands and land area were masked out and filled in with zeros. 
    \item MrCOSTS was applied over four levels with the corresponding window lengths set as $[9,10,30,60]$\,days and singular value decomposition (SVD) ranks as $[4,4,10,12]$, respectively, for each level. 
    \item Spatial modes associated with frequencies smaller than $0.1\,\text{cpd}$ were summed up as the geostrophic SSHa component.
\end{enumerate} 
Note that the spatial smoothing in the second step is not used to extract the geostrophic component but rather to \textit{a priori} remove signals that should definitively not be in geostrophic balance \cite{pedlosky1984equations,savage2017spectral,qiu2018seasonality,Torres2018,Wang_2023}. 
Submesoscale dynamics occupy the space-time continuum between geostrophic dynamics and ITs \cite{zhang2025exploring}, and are not strictly geostrophically balanced \cite{thomas2008submesoscale,mcwilliams2016submesoscale,mcwilliams2019survey,mcwilliams2021oceanic,taylor2022submesoscale}; they are in higher-order balance \cite{McWilliams_2017,uchida2019contribution,chouksey2023comparison,gowthaman2024s,du2025next}. 
In step 5, given that the fast-sampling phase has daily resolution, the data points in time per window correspond to $[9,10,30,60]$ and the SVD ranks need to be smaller than this.

\subsection{Drifters in the Mediterranean Sea}
As an independent data point, we include observations from drifter trajectories in the Mediterranean Sea \cite{medrifters}. 
The dataset comprises of 65 Lagrangian Surface Velocity Program (SVP) drifters that were deployed in the western Mediterranean Sea between March 27$^\text{th}$,~2023 and January 22$^\text{nd}$,~2024.
This amounted in over 27 thousand data points during the SWOT fast-sampling phase.
In this study, we shall restrict our analyses to a subset of this data where the drifters were spatially aligned with SWOT pass No.~3 within a $\pm 6$~hour window (i.e.,~a 12-hour window) of SWOT flyovers, yielding approximately 12{,}700 collocated observations.
Similar to \citeA{muller2019geostrophic}, Ekman contributions were first removed from the drifter velocities after estimating them from ECMWF Reanalysis (ERA5) wind stress reprocessed by Koninklijk Nederlands Meteorologisch Instituut \cite<KNMI;>[]{era5knmi} and using the mean of the empirical model parameters from \citeA{rio2014beyond}. A second-order Butterworth filter with a 48-hour cutoff period was then applied to the drifter velocities. 
The cutoff period was chosen based on rotary frequency power spectra of the drifter velocities so that spectral peaks corresponding to near-inertial oscillations (NIOs) were removed (Section~\ref{sec:svp}).

\section{Results}
\begin{figure}
    \centering
    \includegraphics[width=.9\linewidth]{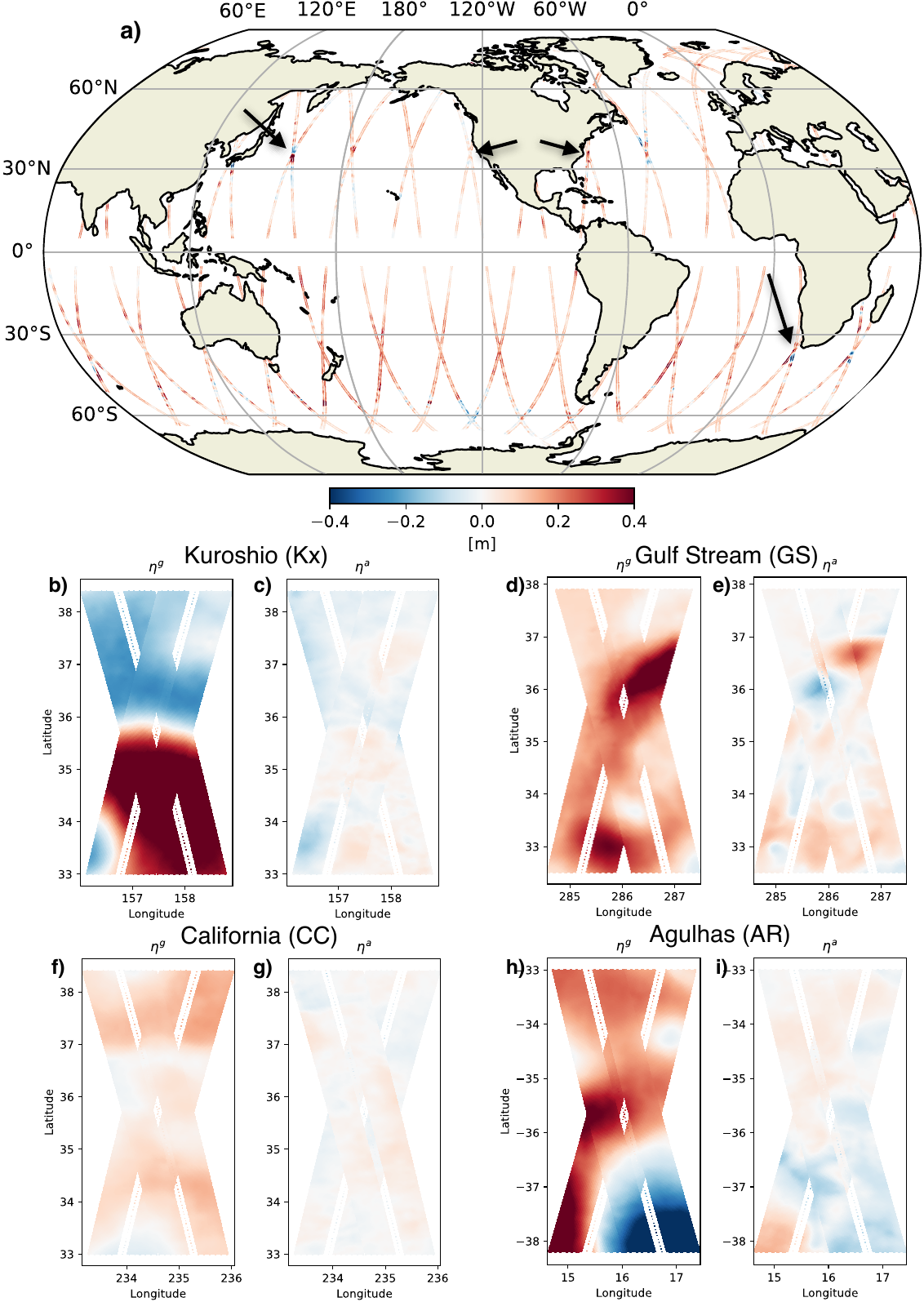}
    \caption{A snapshot of $\eta^g$ on an arbitrary day from the global SWOT Cal/Val orbit and zoomed-in plots of $\eta^g$ and $\eta^a$ at four Xover regions. 
    The four regions, Kx, GS, CC and AR, are indicated by the black arrows in panel (a).}
    \label{fig:ssha}
\end{figure}

We start by documenting a snapshot of geostrophic SSHa, $\eta^g$, in Fig.~\ref{fig:ssha}a.
Hereafter, mathematical notations with the superscript $g$ will be reserved for the geostrophic component filtered through mrCOSTS unless stated otherwise.
The extraction of the geostrophic component naturally yields a residual SSHa field; subtracting the IT and geostrophic component from total SSHa leaves us with the SSHa component associated with ageostrophic eddy dynamics, $\eta^a\,[=\eta - (\eta^\text{IT} + \eta^g)]$. 
$\eta$ is the total SSHa with HRET contributions added back in and $\eta^{\text{IT}}$ is the IT signals from the HYCOM forecast.
The term `eddy' is used here loosely as variability associated with ageostrophic dynamics outside of ITs and isotropic turbulence in the ocean boundary layer.

In the following, we examine the statistical properties at four cross-over (Xover) regions where the ascending and descending passes overlap with each other. This comprises of passes 1 and 16 for the Agulhas Retroflection (AR, between 33$^\circ$S--38.2$^\circ$S), passes 4 and 19 for the Kuroshio extension (Kx, between 33$^\circ$N--38.4$^\circ$N), passes 9 and 22 for the separated Gulf Stream (GS, between 32.5$^\circ$N--37.9$^\circ$N), and passes 13 and 26 for the California Current (CC, between 33$^\circ$N--38.4$^\circ$N) region. 
Dynamically, AR is a region with energetic eddies \cite{jones2023using,carli2023ocean}, and Kx is region with an oceanic jet and energetic eddies but relatively weak ITs \cite{qiu2025fine,yadidya2025predicting}. 
GS is near the Georges Bank and has an oceanic jet with strong eddies and ITs \cite{kelly2016internal,uchida2022cloud,kaur2024seasonal} and was the region also analyzed in \citeA{uchida2025dynamic}.
CC is a relatively quiescent coastal upwelling region \cite{wang2025practical,tchonang2025swot}.
Bottom panels of Fig.~\ref{fig:ssha} zoom in on these regions on $\eta^g$ and $\eta^a$.

\subsection{Frequency and along-track wavenumber power spectra}
We examine the spectral properties of $\eta^{g}$ and $\eta^{a}$ at the four cross-over regions. We take the along-track wavenumber spectra at the cross-track edge of each swath, which minimizes for spatial correlation in the spectral estimates. The Fourier transform in the time dimension are taken over the entire duration of the Cal/Val phase. We apply this procedure for the ascending and descending pass, respectively, and then the frequency-along track wavenumber ($\omega$-$k$) spectral estimates are averaged. 
Namely, the ascending and descending passes are treated as independent observations of SSHa.
Linear trends in space and time were removed and a Hann window was applied prior to taking the Fourier transforms.

\begin{figure}
    \centering
    \includegraphics[width=1.2\linewidth]{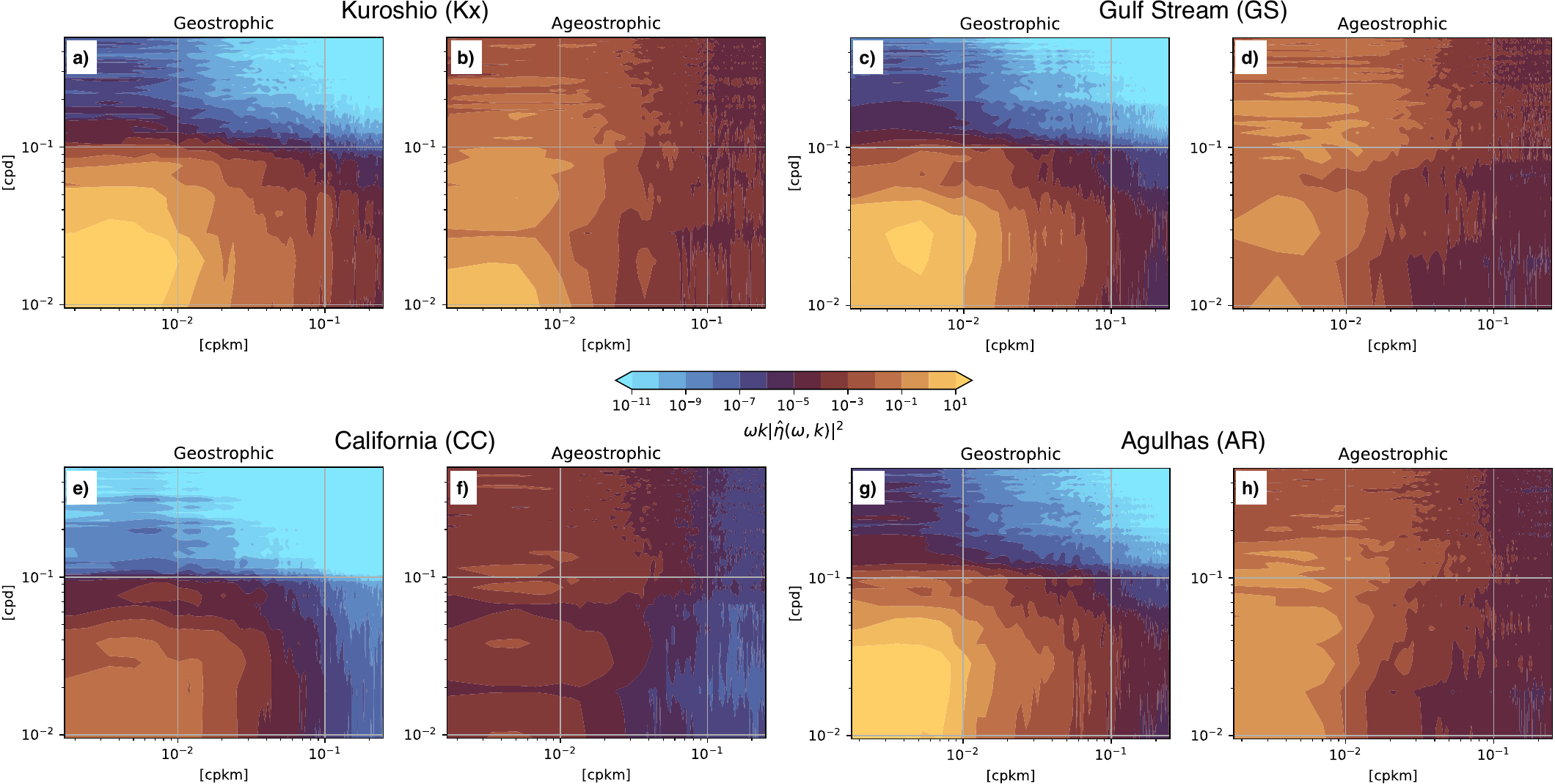}
    \caption{Variance preserving frequency and along-track wavenumber power spectra, $\omega k|\hat{\eta}(\omega,k)|^2~[\text{m}^2\,\text{cpm}^{-1}  \text{cps}^{-1}]$, of $\eta^g$ and $\eta^a$ at four Xover regions. The units on the axes are in cycles per kilometer (cpkm) and cycles per day (cpd).
    }
    \label{fig:omega-k}
\end{figure}

Figure~\ref{fig:omega-k} documents the $\omega$-$k$ power spectra. We observe that the geostrophic spectra peaks at scales larger than 100\,km and slower than 20 days. Notably, there is a clear cutoff of power at time scales shorter than 10 days as expected from the mrCOSTS frequencies (Step 6 in Section~\ref{sec:data}; Fig.~\ref{fig:omega-k}a,c,e,g). The ageostrophic spectra, on the other hand, has a tail that extends into time scales faster than 10 days (Fig.~\ref{fig:omega-k}b,d,f,h). Such distinction in frequency and wavenumber is consistent with the categorization by \citeA<>[their Fig.~3]{Torres2018}.
Amongst the four regions, CC is notably less energetic.

\subsection{Vorticity-strain joint probability distribution}
Relative vorticity and strain rates can be derived from the geostrophic velocity
\begin{equation}
    \zeta^g = v^g_x - u^g_y,\ |\alpha^g| = \sqrt{\left(u^g_x-v^g_y\right)^2 + \left(v^g_x+u^g_y\right)^2}\,,
\end{equation}
where the subscripts are horizontal derivatives.
The geostrophic velocities, $\zeta^g$ and $|\alpha^g|$ were derived from $\eta^g$ following the prescription by \citeA{tranchant2025swot}, which includes centrifugal contributions and is the operational procedure adopted by \citeA<>[Section~2.7 in their handbook, \url{https://www.aviso.altimetry.fr/fileadmin/documents/data/tools/handbook_duacs_SWOT_L3.pdf}]{swot2024}.
\begin{figure}
    \centering
    \includegraphics[width=1.2\linewidth]{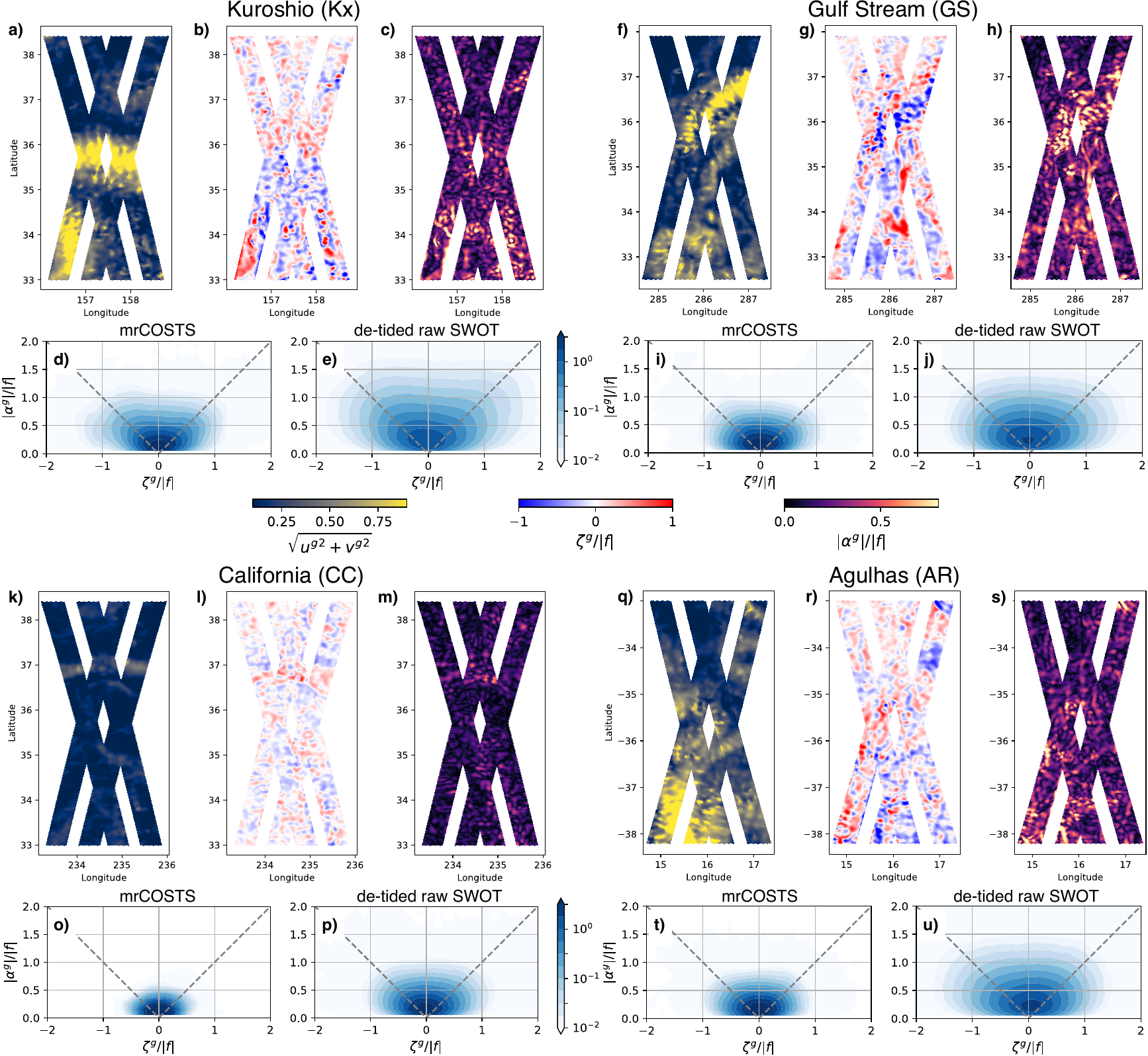}
    \caption{Snapshots of geostrophic speed $|{\bf u}^g|$, relative vorticity $\zeta^g/|f|$ and strain rates $|\alpha^g|/|f|$ pre-processed through mrCOSTS, and joint PDFs of the latter two from four Xover regions.
    Geostrophic speed is shown in the navy-yellow colormap, relative vorticity in blue-red colormap and strain rates in violet-orange colormap.
    The joint PDFs were diagnosed over the three months of $\eta^g$ along with L3$_{\text{HYCOM}}$, i.e.,~the raw SWOT data de-tided by the HYCOM forecast.
    }
    \label{fig:jpdf}
\end{figure}

Figure~\ref{fig:jpdf} displays snapshots of geostrophic speed $\sqrt{{u^g}^2 + {v^g}^2}$, and vorticity and strain rate normalized by the local inertial frequency at the four Xover regions.
The speed in GS and Kx shows the signal of eastward oceanic jets where we would expect them (around 36$^\circ$N; Fig.~\ref{fig:jpdf}a,f).
Again, we see that CC is less energetic compared to the other regions.
In order to make a quantitative description of our geostrophic fields, we diagnose the joint PDF of $\zeta^g/|f|$ and $|\alpha^g|/|f|$; we find that the values processed through mrCOSTS are mostly contained within the acceptable range of smaller than unity ($\zeta^g/|f|<\mathcal{O}(1)$ and $|\alpha^g|/|f|<\mathcal{O}(1)$; Fig.~\ref{fig:jpdf}d,i,o,t).
However, when geostrophic balance is applied to the de-tided raw SWOT fields, L3$_\text{HYCOM}$, the joint PDFs demonstrate a fatter tail towards values larger than unity (Fig.~\ref{fig:jpdf}e,j,p,u).
Namely, the joint PDFs are documenting that geostrophic balance was applied to SSHa signals in L3$_\text{HYCOM}$ that were in fact not in balance.
The diagnostics of L3$_\text{HYCOM}$ are similar to the de-facto procedure of the AVISO L3 product except that ITs were removed using the HYCOM forecast instead of HRET.

\subsection{Evaluation against SVP drifters}
\label{sec:svp}
\begin{figure}
    \centering
    \includegraphics[width=\linewidth]{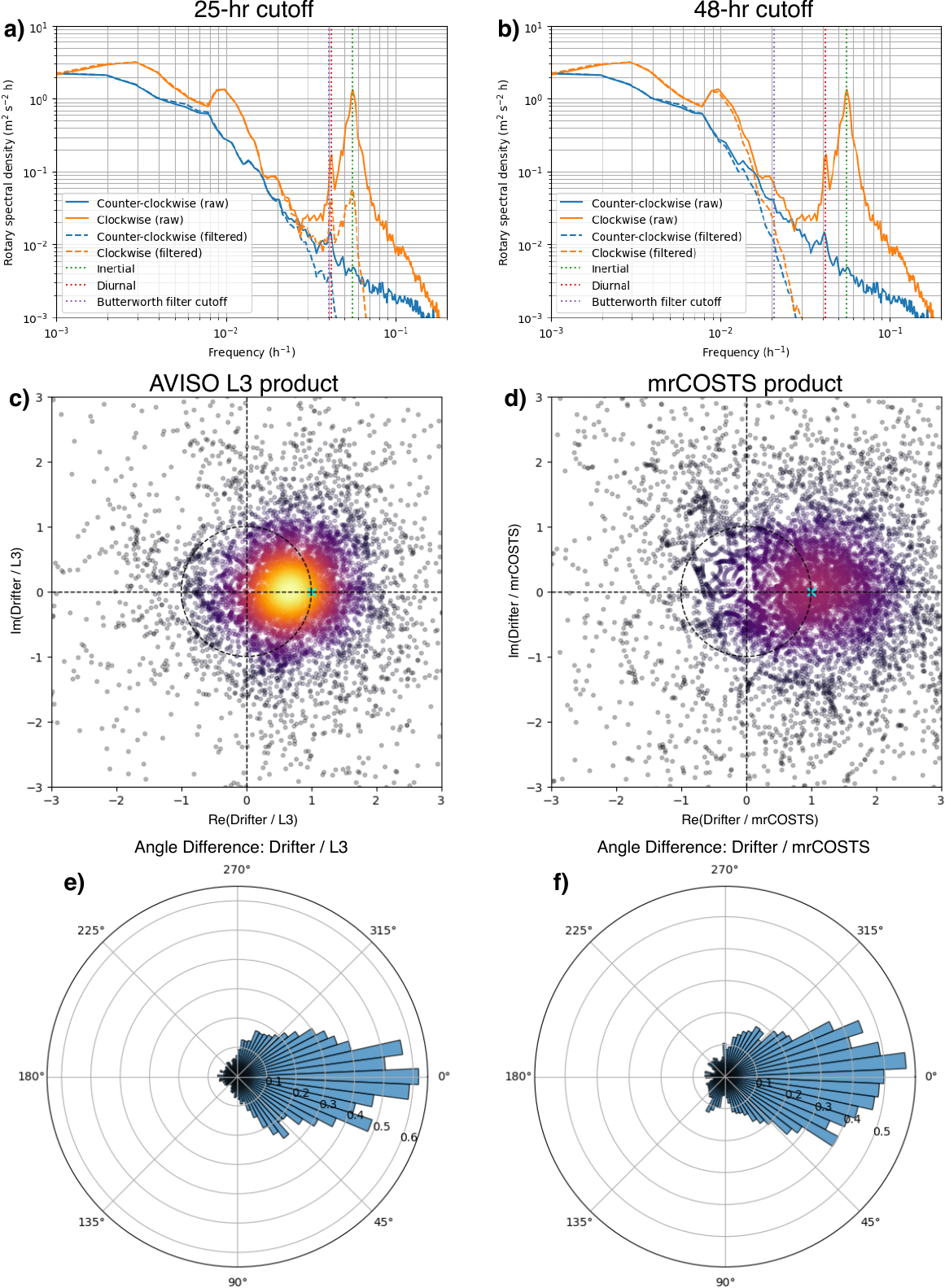}
    \caption{Rotary frequency spectra of the SVP drifter velocities and quantification of how velocity estimates from the L3$_{\text{HRET}}$ and mrCOSTS product agree with them. The unfiltered spectra are shown in solid curves whereas the spectra of velocities low passed with the Butterworth filter are in dashed curves (a,b). The vertical dotted lines correspond to the inertial, diurnal and filter-cutoff frequency.
    Polar histograms between the drifter velocities and L3$_{\text{HRET}}$ and mrCOSTS estimates (c--f). A perfect alignment would collapse the histograms onto the cyan cross and $0^\circ$ in angle.}
    \label{fig:drifter}
\end{figure}

Figure~\ref{fig:drifter}a,b exhibit the clockwise and counter-clockwise rotary frequency spectra of the drifter velocities. 
As expected, we find spectral peaks of NIOs and ITs in the clockwise spectra (solid orange curves). The signal persists under a low-pass Butterworth filtering with a 25-hour cutoff period and is only sufficiently removed when the period is extended to 48 hours (dashed orange curves).
A shift in frequencies of NIOs in the presence of geostrophic flows is a well documented phenomenon and topic of active research \cite{young1997propagation,elipot2010modification,conn2024interpreting}.

We quantify the agreement in magnitude and phase of velocities between the drifters and the default L3 product de-tided with HRET \cite<>[hereon referred to as the L3$_{\text{HRET}}$ product for brevity]{swot2024} and mrCOSTS product using polar histograms of the ratio between each two (represented as complex numbers; Fig.~\ref{fig:drifter}c,d).
The polar histogram would collapse onto the cyan cross upon perfect agreement between the two variables, i.e., $r(\cos{\phi}+i\sin{\phi}) = 1$ where $r=1$ and $\phi = 0$. 
The distance from the unit radius shows the misalignment in magnitude (viz.,~inside the unit circle indicates that the drifter speed is smaller and vice versa) and angle along the unit circle describes the disagreement in orientation.
The L3$_{\text{HRET}}$ speed is too large (the histogram peaks within the unit circle; Fig.~\ref{fig:drifter}c), which is perhaps not surprising since gradients are taken for SSHa fields that are not strictly in geostrophic balance.
We find that both histograms are largely symmetrical about the zero crossing in imaginary number (Fig.~\ref{fig:drifter}c,d), but the L3$_{\text{HRET}}$ product appear to be clustered more tightly around it than mrCOSTS. 
This implies that the former is slightly better aligned with the drifters, also corroborated by the fact that the difference in angles for the L3$_{\text{HRET}}$ product are slightly more centered around zero than mrCOSTS (Fig.~\ref{fig:drifter}e,f).

The spread in the mrCOSTS polar histograms (Fig.~\ref{fig:drifter}d,f) is attributable to: i) differences in the temporal alignment and filtering (10-day Eulerian low-pass filtering versus 48-hour Lagrangian low-pass filtering), ii) potential errors in the estimation of Ekman currents and/or calibration that goes into SWOT prior to product release, and iii) factors other than geostrophy, such as submesoscale dynamics and fronts, affecting the drifters.
Notably, there is a spectral peak around $10^{-2}$\,cph ($\sim 4$ days) in the clockwise rotary spectra.
Nonetheless, it is encouraging that the overestimation in speed we see in the L3$_{\text{HRET}}$ product is alleviated by processing the data through mrCOSTS.
Namely, the center of weight of the polar histogram comes closer to the cyan cross for mrCOSTS than L3$_{\text{HRET}}$.

\section{Discussion and Conclusions}
It is plausible, and likely, that machine learning methods will be able to capture velocities from sea-surface information associated with Rossby numbers on the order of unity and larger \cite<e.g.,>[]{sinha2021estimating,martin2025generative,le2025vardyn,zhou2025machine,ding2025phys,lenain2025unprecedented,wang2025disentangling,lyu2024multi}.
They are also showing promise to focus on the IT field \cite{liu2025wide} or enhance the identification of submesoscale features from SWOT \cite{cutolo2025simulation}.
We would like to argue, however, that a robust isolation of geostrophy sets the base line for extracting velocity information from altimetry observations.

We have applied mrCOSTS, a DMD-based method, to the global SWOT fast-sampling phase to identify the geostrophic SSHa component, $\eta^g$.
The novelty here is that despite the relatively coarse sampling frequency of once per day, we were able to utilize the temporal information along with spatial in achieving the decomposition. 
In the practical sense, our results demonstrate that mrCOSTS overcomes the issue of `spectral leakage' due to temporal aliasing \cite{xu2005antileakage}, and produces dynamically viable spatial modes at frequencies distinctly lower than the Nyquist frequency.
The entire fast-sampling phase was processed and we publicly distribute the geostrophic fields as a data product \cite<upon acceptance of the manuscript;>[]{geoscalval}.

We have documented the $\omega$-$k$ power spectra of SSHa, joint PDF of vorticity and strain rates, and a comparison to drifters in the western Mediterranean Sea as the metrics to assess our geostrophic fields. 
Our velocity estimates do not perfectly align with the drifters, but their agreement is at least as good as the default L3$_{\text{HRET}}$ product if not better, and have the following desirable characteristics:
The $\omega$-$k$ spectra document that $\eta^{g}$ evolves on time scales slower than $\mathcal{O}(10~\text{days})$ and spatial scales larger than $\mathcal{O}(100\,\text{km})$. The ageostrophic component, $\eta^a$, displays higher power into larger frequencies and wavenumber.
The joint PDFs show that Rossby numbers associated with $\eta^g$ are generally smaller than unity, $\zeta^g/|f| < \mathcal{O}(1)$.
We are unable, however, to capture the skewness towards cyclonic features expected from geostrophic eddies \cite{shcherbina2013statistics,balwada2021vertical,jones2023using}.
\citeA<>[their Fig.~7]{uchida2025dynamic} attributed this to the limited amount of Cal/Val data over the duration of three months with 24-hourly resolution (i.e.,~$\sim100$ repetitions of each pass) from which mrCOSTS discovered the modes.
Nonetheless, the joint PDF from $\eta^g$ is better than when geostrophy is applied immediately to de-tided SWOT data;
this supports our opening remark that spatial filtering of SSHa alone is insufficient to isolate geostrophy.

Looking forward, our dataset provides an avenue to diagnose the geostrophic kinetic energy (KE) cascade \cite<cf.,>[]{wang2025cross,qiu2025fine} and vertical velocity via the quasi-geostrophic (QG) or surface QG Omega equation \cite<cf.,>[]{qiu2020reconstructing,barcelo2021fine,carli2024surface}; horizontal velocity in the QG Omega equations should be in geostrophic balance \cite{hoskins2003omega}.
Namely, the high frequency and wavenumber variability we observe in the $\omega$-$k$ power spectra of $\eta^a$ will contaminate the estimation of KE spectral cascade and QG vertical velocity under assuming geostrophy for $\eta - \eta^{\text{IT}}$. The variability appears to be non-negligible even in the relatively quiescent region of CC. Error quantification stemming from this is beyond the scope of this study and will be left for consideration elsewhere.

Regarding the 21-day-repeat orbit of the SWOT science phase, there has been an accumulation of two-and-a-half years of data as of writing of this manuscript. However, this amounts to roughly 50 repeated swaths per pass, which is still insufficient for mrCOSTS to robustly perform. This is because DMD-based methods require a timeseries of the data to `learn' the temporal information \cite{Dylewsky_2019,cardinale2025spectral}.
Such limitations stemming from data quality and quantity are not unique to DMD but rather universal to data-driven methods \cite<e.g.,>[]{budach2022effects,chen2023machine,smith2023temporal,mojgani2024interpretable}.
It remains to be seen with the on-going SWOT operation whether DMD-based methods will be able to produce meaningful results for the science phase.

\section*{Open Research Section}
Level 3 SWOT data (2-km, version~2.0.1) were accessed from \citeA<>[\url{https://www.aviso.altimetry.fr/en/data/products/sea-surface-height-products/global/swot-l3-ocean-products.html}]{swot2024}.
The altimeter products were produced by Ssalto/Duacs and distributed by $\text{AVISO}+$, with support from CMEMS (\url{https://www.aviso.altimetry.fr}).
IT component of HYCOM SSHa (version~1.0) were obtained from \citeA<>[\url{https://doi.org/10.7910/DVN/QQUQNZ}]{DVN/CEY4PQ_2025}.
SVP drifter data are available from \citeA<>[\url{https://www.seanoe.org/data/00896/100828/}]{medrifters}.
ERA5 data was accessed via \url{https://doi.org/10.24381/cds.adbb2d47}.
Spatial filtering was taken using the {\tt gcm-filters} Python package \cite{grooms2021diffusion,loose2022gcm}.
We graciously thank the developers of {\tt PyDMD}, an open-source Python package used to apply mrCOSTS \cite<>[\url{https://github.com/PyDMD/PyDMD/tree/master/tutorials/tutorial20}]{demo2018pydmd,ichinaga2024pydmd}.
We acknowledge \citeA<>[\url{https://github.com/treden/SwotDiag.git}]{tranchant2025swot} for making their code publicly available.
Fourier transforms were taken using the {\tt xrft} Python package \cite{xrft2021}.
Joint PDFs were computed using the {\tt xhistogram} Python package \cite{xhist2021}.
Jupyter notebooks used for mrCOSTS analyses are available via Github \cite<>[a DOI will be added and the geostrophic data product will be publicly distributed upon acceptance of the manuscript]{geoscalval}.

\section*{Conflict of Interest declaration}
The statements, findings, conclusions, and recommendations are those of the author(s) and do not necessarily reflect the views of the U.S. Navy or the Department of Defense.

\acknowledgments
T.~Uchida acknowledges support from the National Science Foundation (NSF) grants OCE-2123632 and OCE-1941963 during his time in the U.S., and the Moscow Institute of Physics and Technology (MIPT) Development Program (Priority-2030) upon moving to Russia.
B.~Yadidya, B.~Arbic and J.~Shriver acknowledge funding support from the Office of Naval Research (ONR) grants N00017-22-1-2576 (BY, BA) and N000142WX01587-1 (JS), which are components of the Global Internal Waves project of the National Oceanographic Partnership Program (\url{https://nopp-giw.ucsd.edu/}). 
B.~Arbic and J.~Shriver also acknowledge funding from the National Aeronautics and Space Administration (NASA) grants 80NSSC20K1135 and 80NSSC24K1649.
T.~Uchida personally thanks Mikhail~Borisov and Alexander~Gavrikov for maintaining the Neptun cluster at MIPT and Kubrick cluster at IORAS, respectively, on which the analyses were executed.
This work is a contribution to the Working Group on ocean fine-scale processes (WG57) under the North Pacific Marine Science Organization (PICES; \url{https://meetings.pices.int/members/working-groups/wg57}).

%
%

\bibliography{ref}

%
%
%
%
%

\end{document}